\documentclass[aps,prl,showpacs,reprint,amsmath,amssymb]{revtex4-1}

\usepackage{siunitx}
\usepackage{graphicx}
\usepackage{dcolumn}
\usepackage{bm}
\usepackage{hyperref}
\begin{document}


\title{Interfacing transitions of different alkali atoms and telecom bands using \\one narrowband photon pair source}
\pdfoutput=1
\author{Gerhard Schunk$^{1,2,3}$}
\author{Ulrich Vogl$^{1,2}$}
\author{Dmitry V. Strekalov$^{1,2}$}
\author{Michael F\"{o}rtsch$^{1,2,3}$}
\author{Florian Sedlmeir$^{1,2,3}$}
\author{Harald G. L. Schwefel$^{1,2,4}$}
\author{Manuela G\"{o}belt$^{1}$}
\author{Silke Christiansen$^{1,5}$}
\author{Gerd Leuchs$^{1,2}$}
\author{Christoph Marquardt$^{1,2,6}$}
\affiliation{$^{1}$Max Planck Institute for the Science of Light, G\"{u}nther-Scharowsky-Stra\ss e 1/Building 24, 90158 Erlangen, Germany}              
\affiliation{$^{2}$Institute for Optics, Information and Photonics, University Erlangen-N\"{u}rnberg, Staudtstr.7/B2, 90158 Erlangen, Germany}
\affiliation{$^{3}$SAOT, School in Advanced Optical Technologies, Paul-Gordan-Str. 6, 91052 Erlangen, Germany}
\affiliation{$^{4}$Department of Physics, University of Otago, 730 Cumberland Street, Dunedin, New Zealand}
\affiliation{$^{5}$Helmholtz Centre Berlin for Materials and Energy, Hahn-Meitner Platz 1, 14109 Berlin, Germany}
\affiliation{$^{6}$Department of Physics, Technical University of Denmark, Building 309, 2800 Lyngby, Denmark}

\begin{abstract}
Quantum information technology strongly relies on coupling of optical photons with narrowband quantum systems, such as quantum dots, color centers, and atomic systems. This coupling requires matching the optical wavelength and bandwidth to the desired system, which presents a considerable problem for most available sources of quantum light. Here we demonstrate coupling of alkali dipole transitions with a tunable source of photon pairs. Our source is based on spontaneous parametric down-conversion in a triply-resonant whispering-gallery mode resonator. For this, we have developed novel wavelength tuning mechanisms, which allow for a coarse tuning to either cesium or rubidium wavelength with subsequent continuous fine-tuning to the desired transition. As a demonstration of the functionality of the source, we performed a heralded single photon measurement of the atomic decay. We present a major advance in controlling the spontaneous down-conversion process, which makes our bright source of single photons now compatible with a plethora of narrow-band resonant systems.
\end{abstract}

\maketitle

\section{Introduction}
Photon pairs and heralded single photons are key prerequisites for many schemes in quantum information processing \cite{Kimble2008,Knill2001}. Making their central frequency and bandwidth compatible to resonances of other physical systems enables a variety of applications, such as efficient quantum memories \cite{Lvovsky2009,Simon2010}, photon-phonon interactions \cite{Aspelmeyer2014}, and coupling single photons or squeezed light with single atoms \cite{Specht2011} or atomic ensembles \cite{Hald1999,Hammerer2005}. It also provides the technology for quantum repeater schemes \cite{Sangouard2011} that are proposed to enhance long-distance quantum communication \cite{Duan2001,Gisin1991}.

Presently available single photon sources \cite{Eisaman2011} do not provide quantum-mechanically pure single-mode optical states with a high tunability in wavelength and bandwidth. Previously we reported a source based on spontaneous parametric down-conversion (SPDC) in an optical resonator operating in a single mode regime \cite{Michael2014}, and providing tunable bandwidth \cite{Michael2013} compatible with atomic transitions. Here we demonstrate tuning of the signal photon wavelength to the rubidium and cesium transitions while the idler photon matches a telecom wavelength. 

Spontaneous nonlinear conversion in $\chi^{(2)}$ \cite{Ou1999,Fasel2004,URen2004,Fedrizzi2007,Scholz2009,Wolfgramm2011,Michael2013,Harder2013,Fekete2013} and $\chi^{(3)}$ materials \cite{Soller2010,Iang2015} offers correlated photon pairs with widely tunable wavelength in contrast to many other sources of single photons such as single atoms \cite{Kuhn2002,Maiwald2012,Utikal2014}, single molecules \citep{Lounis2000,Siyushev2014}, solid-state quantum dots \cite{Kurtsiefer2000,Michler2000,Solomon2002}, or four-wave mixing in atomic ensembles \cite{Kuzmich2003,Srivathsan2013,Hengwang2014}. The challenge is to make spontaneous nonlinear conversion both continuously tunable and narrowband to be compatible with quantum memories.

To achieve this we use a triply-resonant cavity. This scheme increases the efficiency of spontaneous nonlinear conversion processes and provides the narrow bandwidth of the cavity in use. Continuous tuning of the output bandwidth in a triply-resonant whispering-gallery mode resonator is possible \cite{Michael2013} by changing the out-coupling rates. Recently, we showed single mode operation \cite{Michael2014} in such a system based on the highly restrictive phase-matching conditions of the triply-resonant cavity. 

In this work, we demonstrate novel tuning mechanisms to generate photon pairs matching two different narrowband atomic transitions in the near infrared and telecom bands. We use the tunable signal photons (790 - \SI{1064}{nm}) for addressing the atomic transitions and the corresponding idler photons (1064 - \SI{1630}{nm}) for heralding at telecom wavelengths. For precise tuning of the signal frequency to the Cs D1 line at \SI{895}{nm} and the Rb D1 line at \SI{795}{nm}, we introduce a set of tuning mechanisms based on mode analysis \cite{Schunk2014a} of the WGMR spectrum at the pump wavelength. Relying solely on temperature tuning one is limited to discrete solutions of the phase-matching conditions. We show continuous tuning of the signal frequency across the Doppler-broadened absorption line of a Cs D1 hyperfine transition by manipulating the evanescent fields of pump, signal, and idler with a movable dielectric substrate. This continuous frequency-tuning with MHz-precision allows for heralded single photon spectroscopy which we demonstrate at this Cs D1 hyperfine transition.
\section{Experimental scheme}
\begin{figure}[htbp]
\centering
\includegraphics[scale=1]{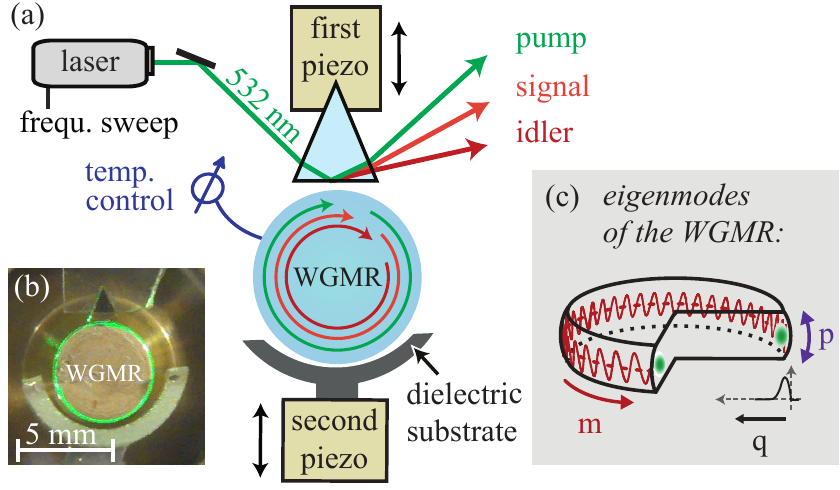}
\caption{Experimental setup. (a) Type-I parametric down-conversion in a z-cut whispering-gallery mode resonator (WGMR) is performed with a pump beam at \SI{532}{\nm} coupling via the diamond prism on top. A second coupling port can be used for dielectric tuning. Alternatively, the dielectric substrate can be replaced with a second diamond prism for mode analysis \cite{Schunk2014a}. (b) Top view of the WGMR. (c) Illustration of the eigenmodes of WGMRs. These eigenmodes are defined by the azimuthal mode number m, the radial mode number q and the angular mode number p.
}
\label{fig:setup1}
\end{figure}

Our source of single photons \cite{Michael2013} and squeezed light \cite{Furst2011} is based on natural phase-matching \cite{Furst2010natural} in a whispering-gallery mode resonator with radius $R=\SI{2.5}{\mm}$ manufactured from a congruent \SI{5.8}{\percent} MgO-doped lithium niobate wafer (see Fig. \ref{fig:setup1}). 
The temperature of the WGMR is controlled at the millikelvin scale with a proportional-integral-derivative (PID) lock. As a pump source we use a frequency-doubled Nd:YAG laser at \SI{532}{\nm} in vertical polarization, which is coupled to the WGMR via a diamond prism (see Fig.~\ref{fig:setup1}). We use a second piezo actuator for moving a glass substrate, which is coated with zinc oxide. This allows for continuous tuning of the resonance frequencies of the eigenmodes. Alternatively, the dielectric substrate can be replaced with a second diamond prism for an analysis of the whispering-gallery modes \cite{Schunk2014a}.

The solution of Maxwell's equations for WGMRs with azimuthally symmetric geometry \cite{Breunig2013a,Demchenko2013} is characterized by three mode numbers (see Fig.\ref{fig:setup1}(c)): the azimuthal mode number $\textrm{m}\gg1$, the radial mode numbers $\textrm{q} \geq 1$, and the angular mode number $\textrm{p}\geq 0$. In the experiment we analyzed the mode spectrum to identify the various eigenmodes of the pump including the fundamental mode with $\textrm{m}_\textrm{p}\approx 64\,900,\textrm{q}_\textrm{p}=1$, and $\textrm{p}_\textrm{p}=0 $ at the temperature $\textrm{T}=140\,^\circ$C. The loaded quality factor at critical coupling and free spectral range (FSR) of the fundamental mode were measured to be $\textrm{Q}=1.6 \times 10^{7}$, and $\textrm{FSR}=\SI{7.8}{\GHz}$, respectively.

\section{Tuning mechanisms for parametric down-conversion in a triply-resonant WGMR}
\subsection*{Temperature tuning}
\begin{figure}[t!!!]
\centering
\includegraphics[scale=1]{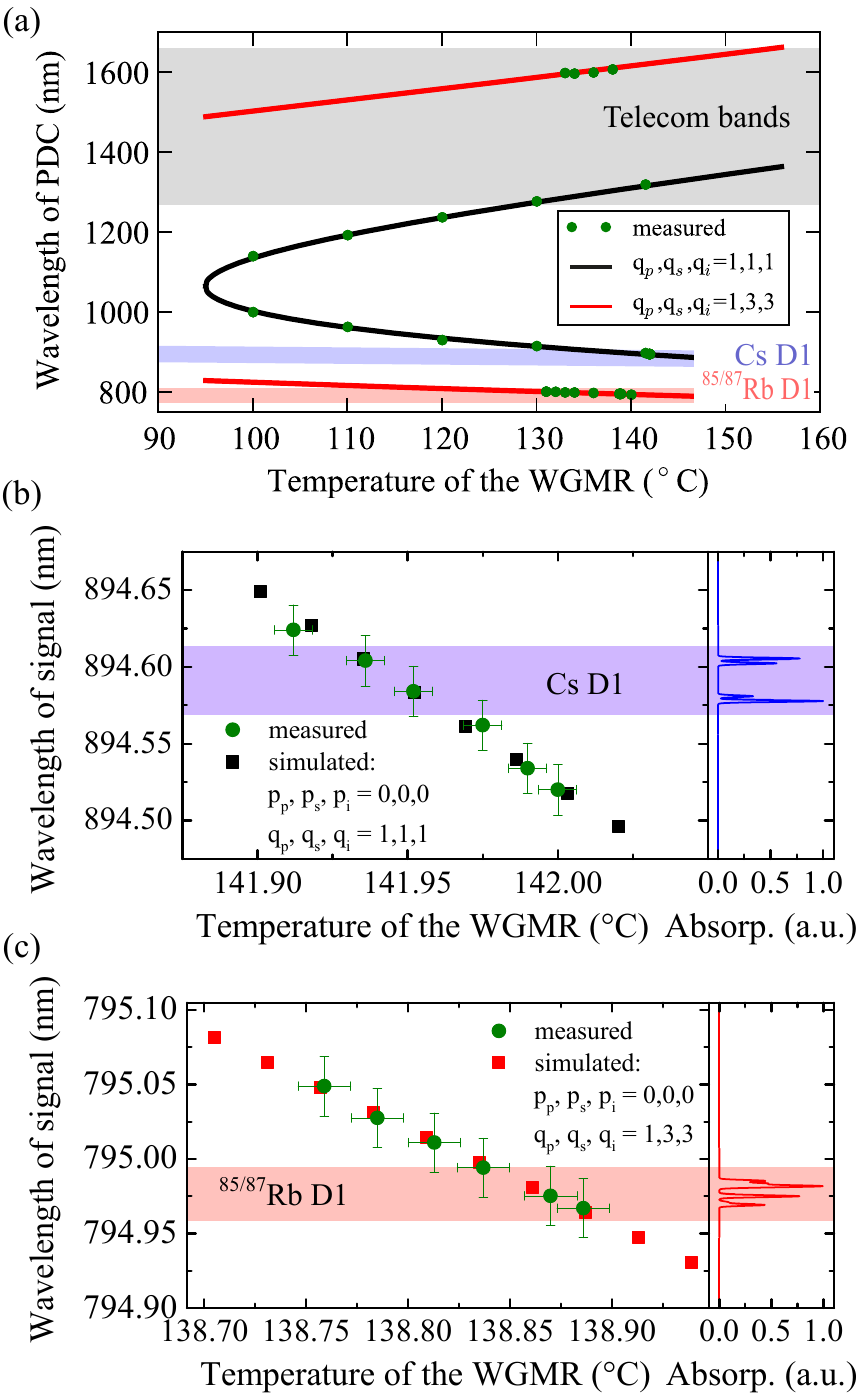}
\caption{Temperature tuning of the parametric down-conversion process in our triply-resonant optical parametric oscillator above the pump threshold. The generated signal and idler beams connect various alkali atoms to telecom wavelengths. (a) Emission wavelengths from \SI{790}{\nm} to \SI{1630}{\nm} are measured for a fundamental pump mode with $\textrm{m}_\textrm{p}\gg1,\textrm{p}_\textrm{p}=0$, and $\textrm{q}_\textrm{p}=1$ emitting into various signal and idler mode combinations. We show the branches relevant for tuning to the D1 lines of Cs and $^{85/87}$Rb, respectively. (b) Temperature fine tuning of the fundamental conversion channel $\textrm{q}_\textrm{p},\textrm{q}_\textrm{s},\textrm{q}_\textrm{i}=1,1,1$ around the Cs D1 line shows the stepwise tuning behavior keeping the the azimuthal mode number $\textrm{m}_\textrm{p}$ of the pump fixed. (c) The same stepwise tuning behavior is found for the non-fundamental conversion channel $\textrm{q}_\textrm{p},\textrm{q}_\textrm{s},\textrm{q}_\textrm{i}=1,3,3$ around the  $^{85/87}$Rb D1 line. The right panels show calculated absorption profiles for Cs and $^{85/87}$Rb, respectively \cite{Steck2008all}. The shown uncertainties result from the limited resolution of the optical spectrum analyzer (AQ6370C, Yokogawa).}
\label{fig:temperaturetuning}
\end{figure}
The most efficient conversion channel \cite{Furst2010} of our parametric down-conversion (PDC) process converts light from the extraordinarily polarized fundamental pump mode to the ordinarily polarized fundamental signal and idler modes with $\textrm{m}_\textrm{s}\gg1,\textrm{q}_\textrm{s}=1,\textrm{p}_\textrm{s}=0 $  and $\textrm{m}_\textrm{i},\textrm{q}_\textrm{i}=1,\textrm{p}_\textrm{i}=0 $, respectively. Fig.~\ref{fig:temperaturetuning}(a) shows the measured PDC frequencies of this triply-resonant optical parametric oscillator (OPO) pumped above threshold in a temperature range from \SIrange[range-units = single]{100}{140}{\degreeCelsius} along with a numerical simulation of conversion channels for higher-order radial PDC modes. Depending on the temperature, the mode of the fundamental pump defined by $\textrm{m}_\textrm{s}$ is selected  for a fixed pump laser wavelength. In this graph we omit non-relevant conversion channels for clarity, which may be accessed by considering modes with different p and q values \cite{Werner2012,Michael2014}.

Fig.~\ref{fig:temperaturetuning}(b) and (c) show the stepwise tuning behavior for the fundamental conversion channel around the Cs and Rb D1 lines, respectively. These steps originate from the interplay between the discrete eigenresonances and the phase-matching requirements. Phase-matching in WGMRs \cite{Kozyreff2008} is mainly characterized by:
\begin{subequations}
	\begin{align}
	\nu^\textrm{ }_\textrm{p} (\textrm{T},\textrm{d}) &= \nu^\textrm{ }_\textrm{s} (\textrm{T},\textrm{d}) + \nu^\textrm{ }_\textrm{i} (\textrm{T},\textrm{d}) \,, 
	\label{eq:energycons} \\
	\textrm{m}_\textrm{p} &= \textrm{m}_\textrm{s} + \textrm{m}_\textrm{i} \,,
	\label{eq:momentumcons}
	\end{align}
	\label{eq:phasematch}
\end{subequations}
representing energy and momentum conservation, respectively. 
The resonance frequencies $\nu^{\textrm{ }}_{\textrm{p,s,i} } (\textrm{T},\textrm{d}) $ depend on the temperature T and the distance d between the WGMR and the dielectric substrate controlled with the second piezo actuator. Eq.~\ref{eq:momentumcons} is equivalent to the conservation of angular momentum for electronic transitions in atomic physics and explains the stepwise tuning behavior in our case. For small temperature changes we can assume $\nu^{\textrm{ }}_{\textrm{p,s,i} } (\textrm{T}) $ to be linear functions in temperature. If Eq.~\ref{eq:phasematch}a,b are fulfilled for a fixed triplet of modes with $\textrm{m}_\textrm{p,s,i},\textrm{q}_\textrm{p,s,i},$ and $\textrm{p}_\textrm{p,s,i}$ at a given temperature T$_0$, then phase-matching for this mode combination is typically allowed only in a narrow interval, i.e. the phase-matching bandwidth around T$_0$. The magnitude of the phase-matching bandwidth is determined by the bandwidths of the three modes and the different slopes of $\nu^{\textrm{ }}_{\textrm{p,s,i} } (\textrm{T},\textrm{d})$. However,  phase-matching for:
\begin{align}
	\textrm{m}_\textrm{p},\textrm{m}_\textrm{s},\textrm{m}_\textrm{i} \rightarrow \textrm{m}_\textrm{p},\textrm{m}_\textrm{s} \pm 1,\textrm{m}_\textrm{i} \mp 1  \,,
	\label{eq:PDCkeeppump}
\end{align}
is allowed according to Eq.~\ref{eq:phasematch}a,b at a temperature T$\pm\Delta$T keeping the same pump mode. The temperature steps $\Delta$T are determined by the different slopes of $\nu^\textrm{ }_{\textrm{p,s,i} }(\textrm{T})$ and the difference in the respective FSR of signal and idler.
\begin{figure}[ht!!!]
\centering
\includegraphics[scale=1]{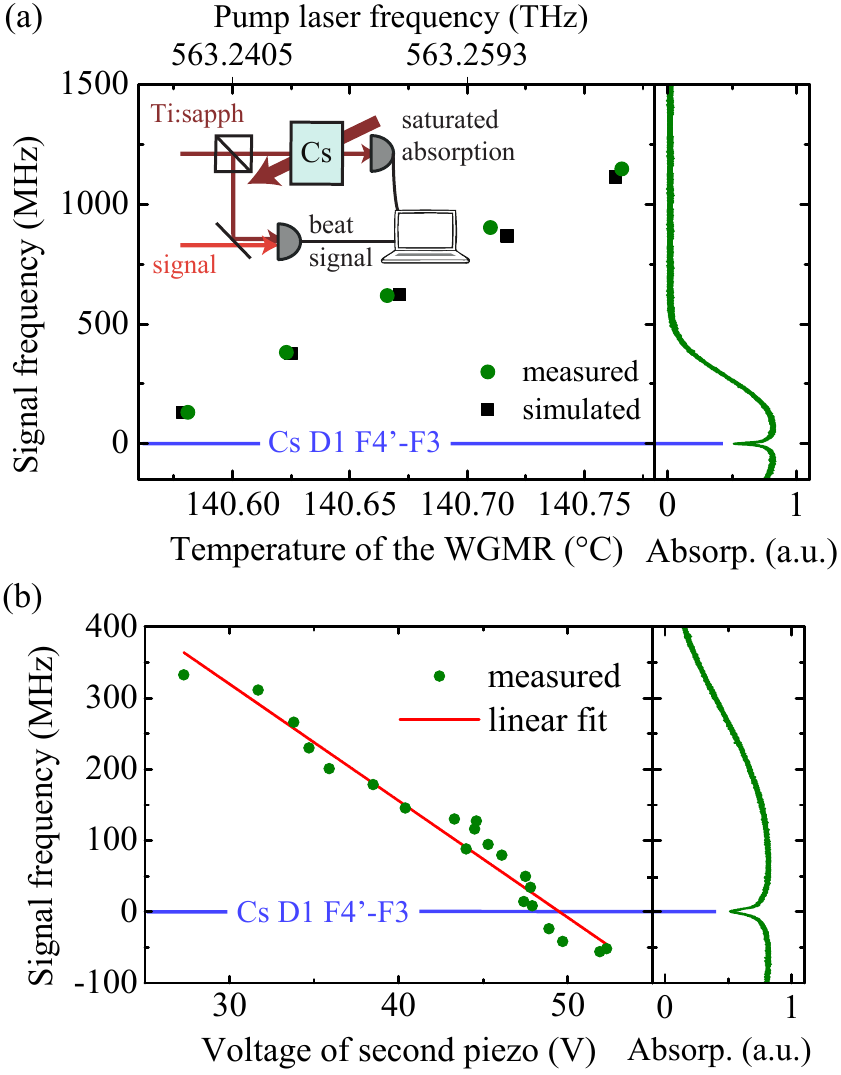}
\caption{Frequency tuning of the parametric down-conversion process measured via a beat signal between the signal beam and the Ti:sapphire reference laser in a saturated absorption configuration. The right panels show the measured Ti:sapphire absorption at the Cs D1 F4'-F3 line, including the Doppler-free dip on resonance. (a) A change in the azimuthal mode number $\textrm{m}_\textrm{p}$ of the pump mode accompanied by a temperature adjustment results in an average change in the signal emission frequency by $\Delta \nu^\textrm{ }_\textrm{s}=\SI{254}{\MHz}$. (b) Tuning across the Doppler-broadened line is achieved by shifting the resonance frequencies of the pump and the parametric modes with a movable dielectric substrate in the evanescent fields (we estimate the distance change per applied voltage on the actuator to be \SI{15}{\nm\per\V}).}
\label{fig:BeatMeas}
\end{figure}
\subsection*{Selection of the pump mode}
For measuring the absolute signal frequency on the MHz scale, we beat the signal with a Ti:sapphire reference laser. This reference laser scans over the approximately $\SI{500}{\MHz}$ wide Doppler-broadened Cs D1 line in a saturated absorption configuration. The beat signal then only appears when the frequency difference between the reference laser and the signal is within the detector bandwidth (PDA36A from Thorlabs with \SI{11}{\kHz} bandwidth). Notably, above the threshold the bandwidth of parametric light \cite{Debuisschert1993} is determined by the pump bandwidth in contrast to the sub-threshold operation when it is determined by the resonator linewidth (see Fig.~\ref{fig:spdcabs} and \ref{fig:spdc}).The atomic line data for Cs and $^{85/87}$Rb were taken from \cite{Steck2008all}. By changing the pump mode azimuthal mode number as:
\begin{align}
	\textrm{m}_\textrm{p},\textrm{m}_\textrm{s},\textrm{m}_\textrm{i} \rightarrow \textrm{m}_\textrm{p} \pm 1,\textrm{m}_\textrm{s},\textrm{m}_\textrm{i} \mp 1  \,,
	\label{eq:PDCkeepsignal}
\end{align}
we achieve an overall tuning of the signal frequency of approximately \SI{1}{\GHz} in steps of \SI{254}{\MHz} presented in Fig.~\ref{fig:BeatMeas}(a). Notably, the mode numbers $\textrm{m}_\textrm{s}\gg1,\textrm{q}_\textrm{s}=1$, and $\textrm{p}_\textrm{s}=0$ of the signal do not change in this process and frequency tuning occurs only due to the temperature dependent frequency drift of the same signal mode. Using this approach, the complete interval between two steps of about \SI{0.022}{\nm} (corresponding to \SI{8.2}{\GHz}) shown in Fig.~\ref{fig:temperaturetuning}(b,c) can be covered if an overall tuning of the pump laser over approximately \SI{240}{\GHz} is provided. Furthermore, all conversion channels that are allowed by the phase-matching conditions can be used for tuning to the desired wavelength, which relaxes the requirements on the pump laser tuning. For example, for our tuning to the Rb D1 line we used the non-fundamental conversion channel $\textrm{q}_\textrm{p},\textrm{q}_\textrm{s},\textrm{q}_\textrm{i}=1,3,3$. Here again we measured the beat signal of the generated signal and the Ti:sapphire reference laser at the Doppler-broadened $^{85}$Rb D1 F3'-F2 line.

\subsection*{Perturbation of the evanescent fields}
Triply-resonant OPOs offer efficient PDC and narrowband photon pairs under the constraint of the stepwise tuning behavior discussed in the previous section. An addressing of atomic transitions with arbitrary detuning requires continuous frequency adjustments at the MHz-scale. We achieve continuous tuning of the phase-matching by combining temperature tuning and the effect of a movable dielectric substrate, i.e. a curved glass substrate coated with highly refractive zinc oxide (n$_\textrm{ZnO}$=2.03 at \SI{532}{\nm} \cite{Bond1965}) by atomic layer deposition \cite{George2010}, within the evanescent fields of the WGMR \cite{Teraoka2006}. This does not lead to a reduction of the optical quality factor via out-coupling since the refractive index of the WGMR is higher than the refractive index of the dielectric substrate. As all three modes, the pump, signal and idler, are affected, we adjust the pump laser frequency to the new phase-matching condition. In contrast to changing the mode numbers of the three modes, we change their resonance frequencies by $\delta \nu^\textrm{ }_\textrm{p,s,i}$ while fulfilling the energy conservation requirement given by Eq.~\ref{eq:energycons}. Thereby, the signal frequency $\nu^\textrm{ }_\textrm{s}$ shifts as:
\begin{align}
	  \delta \nu^\textrm{ }_\textrm{s} (\textrm{T},\textrm{d}) &= \delta \nu^\textrm{ }_\textrm{p} (\textrm{T},\textrm{d}) -  \delta \nu^\textrm{ }_	\textrm{i} (\textrm{T},\textrm{d}) \,.
\end{align}
An increase in the voltage of the second piezo actuator, which changes the distance between the dielectric and the WGMR, in combination with temperature adjustments allows for continuous tuning of the signal frequency over approximately \SI{400}{MHz} shown in Fig.~\ref{fig:BeatMeas}(b). This tuning range exceeds the previously reported values of \SI{150}{MHz} achieved via the electro-optical effect of lithium niobate \citep{Michael2013}, which provides an additional method for fast switching.
\section{Coupling single photons to atoms}
In the previous section we show accurate frequency measurements of the generated parametric beams above the OPO threshold using the spectral lines as a reference. In the following we directly show light-atom coupling for the desired single photon states. The single photon regime \cite{Michael2013} in this WGMR-based OPO is reached by reducing the pump power far below the OPO threshold \cite{Ilchenko2003a,Furst2010natural,Beckmann2012} of \SI{18}{\uW} at the temperature T=\SI{141}{\degreeCelsius} with the signal and idler wavelength $\lambda_\textrm{s}=\SI{895}{\nm}$ and $\lambda_\textrm{i}=\SI{1312}{\nm}$, respectively. This threshold was measured for the most efficient conversion channel of the PDC process using fundamental pump, signal, and idler modes with $\textrm{m}_\textrm{p,s,i}\gg1,\textrm{q}_\textrm{p,s,i}=1,$ and $\textrm{p}_\textrm{p,s,i}=0$.
\begin{figure}[ht!!!]
\centering
\includegraphics[scale=1]{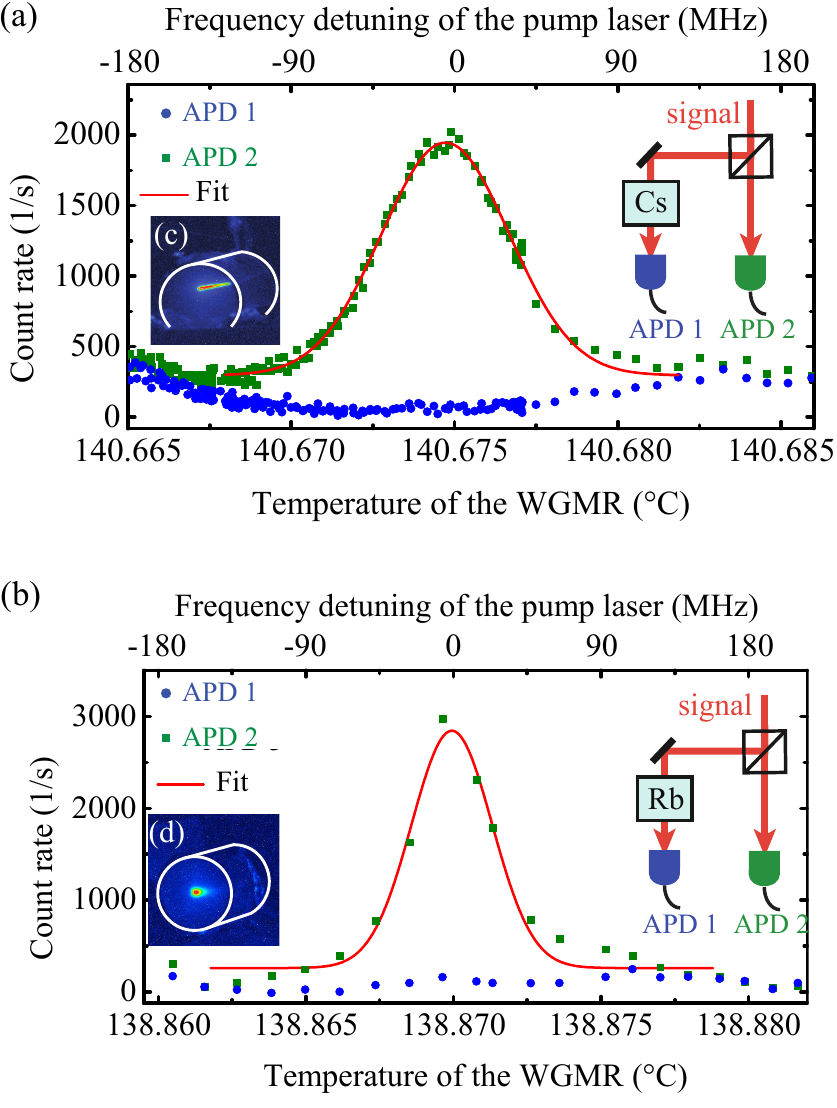}
\caption{Absorption of the signal light by the atoms. (a) In this measurement we decreased the pump power below the OPO threshold to demonstrate resonant single photons. The pump laser frequency was locked to the fundamental mode while the WGMR temperature was continuously changed. We measured a suppression of the count rate by $\SI[multi-part-units = single]{97\pm3}{\percent}$ at the Cs D1 F4'-F3 line in (a) and by $\SI[multi-part-units = single]{98\pm4}{\percent}$ at the $^{85}$Rb D1 F3'-F2 line in (b). The frequency width of the peaks directly gives the respective phase-matching bandwidths of the parametric down-conversion process. Images of bright fluorescence in the Cs cell in (c) and the Rb cell in (d) were taken far above the OPO threshold.}
\label{fig:spdcabs}
\end{figure}

The phase-matching conditions given by Eq.~\ref{eq:phasematch}a,b determine the threshold of the PDC process and also the efficiency of the SPDC process. The possibility to fulfill energy conservation (Eq.~\ref{eq:energycons}) for three given modes is highly dependent on the temperature of the WGMR. In Fig.~\ref{fig:spdcabs} we show the temperature-dependent detuning of the modes from the phase-matching point via the count rates of the signal photons. The idler photons are not considered for this absorption measurement. One half of the signal photon flux is sent through the vapor cells and the other half is directly detected as a reference. APD 1 and APD 2 are Si avalanche photodiode (SPCM CD 3017 from Perkin Elmer). The cell was \SI{5}{\cm} long and at a temperature of \SI{80}{\degreeCelsius}. Interference filters (FB900-25 at $\SI[multi-part-units = single]{900\pm5}{\nm}$ from Thorlabs for the Cs D1 line and ASE-797 at $\SI[multi-part-units = single]{794.979\pm 0.125}{\nm}$ from Ondax for the Rb D1 line) were used to filter out signal modes which are separated by several nanometer from the desired wavelength. 

Below threshold, it is possible to generate photons in multiple PDC conversion channels at the same time. This aspect cannot be deduced from the above threshold measurements presented in Fig.~\ref{fig:BeatMeas} due to the different working principle of PDC above threshold based on self-seeding and the limited detection range of approximately \SI{2}{\GHz} in the beat signal measurement. However, the high absorption values of $\SI[multi-part-units = single]{97\pm3}{\percent}$ for Cs and $\SI[multi-part-units = single]{98\pm4}{\percent}$ for Rb show that all signal photons in this measurement were residing within the approximately \SI{500}{\MHz} wide Doppler-broadening of the alkali atoms. Single-mode operation \cite{Michael2014} without an additional bandpass filter occurs if the phase-matching temperatures of the most efficient conversion channels are separated by more than the phase-matching bandwidth. This can in principle be achieved with an adaption of the radius and the curvature of the WGMR. 
\begin{figure}[ht!!!]
\centering
\includegraphics[scale=1]{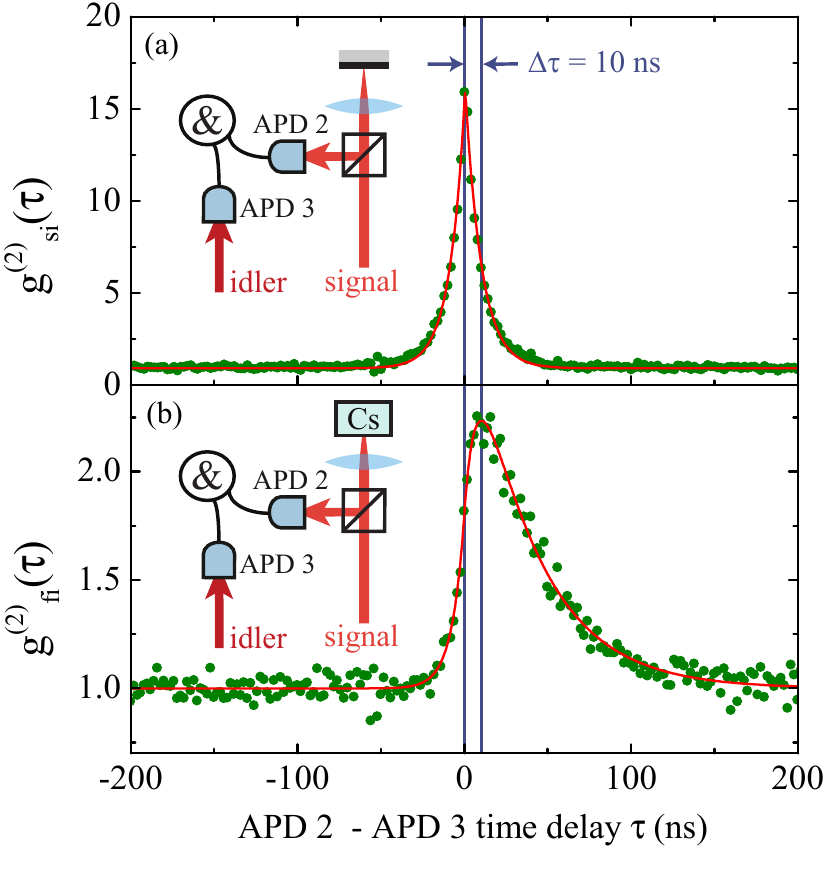}
\caption{Photon-atom coupling. (a) The measured coincidences (\SI{2}{ns} bin width, \SI{120}{\s} integration time) from the directly detected signal and idler photons are fitted with the double exponential function given by Eq.~\ref{eq:directcorr}. This yields the respective decay time of $\tau_\textrm{si}$=9.4\,ns corresponding to a photon bandwidth of $\gamma_\textrm{si}=$\SI{16.9}{MHz}. (b) This measurement of heralded fluorescence (\SI{2}{\ns} bin width, \SI{60}{minutes} integration time) allows for a direct probing of the atomic decay. The time delay of $\Delta\tau =$\SI{10}{\ns} with respect to the reference measurement in (b) is in agreement with the theoretical model given by Eq.~\ref{eq:fluorcorr}.}
\label{fig:spdc}
\end{figure}

Next we investigate temporal correlations of signal and idler at the single photon level while the signal photons are interacting with the cesium atoms. The arrival of signal photons is heralded with idler photons detected with APD 3, which is an InGaAs/InP avalanche photodiode (ID220 from ID Quantique). Fig.~\ref{fig:spdc}(a) shows the measured correlation function $\textrm{g}^{(2)}_\textrm{si}(\tau)$ of the directly detected signal and idler photons with a double-exponential fit given by 
\begin{align}
	\textrm{g}^{(2)}_\textrm{si}(\tau) \propto 
	\frac{1}{2} \cdot   \frac{ e^{- \vert \tau \vert / \tau_\textrm{si}}}{\tau_\textrm{si}} \,,
	\label{eq:directcorr}
\end{align}
at a pump power of \SI{0.39}{\uW}. A comparison of the count rates on each APD with the measured coincidences yields a Klyshko efficiency \cite{Klyshko1980} of \SI{1.5}{\percent} for the idler and \SI{0.9}{\percent} for the signal. Lower limits for the overall losses are only given by the material absorption of lithium niobate, which can be optimized by using crystals of higher optical grade or simply by lowering the decay times by over-coupling the resonator. 

The decay times of signal and idler $\tau_{\textrm{si}}$, i.e. the respective bandwidths $\gamma_{\textrm{si}}= 1 / (2\pi \cdot \tau_{\textrm{si}})$, can be tuned jointly by controlling the distance between coupling prism and WGMR. We measured a bandwidth tuning of approximately $\gamma_{\textrm{si}} = 6-24\,$MHz ($\gamma_{\textrm{si}} = 7-13\,$MHz was shown in Ref. \cite{Michael2013}).

Fig.~\ref{fig:spdc}(b) shows the measured correlation function of directly detected idler photons heralding the photons coming from resonant fluorescence of the Cs D1 F4'-F3 line at a pump power of \SI{3.3}{\uW}. We calculate the correlation function $\textrm{g}^{(2)}_\textrm{fi}(\tau)$ for this heralded fluorescence measurement with a stochastic atomic decay model with an exponential probability distribution characterized with an atomic decay time $\tau_\textrm{f}$:
\begin{align}
	\textrm{g}^{(2)}_\textrm{fi}(\tau) \propto \frac{1}{2} \cdot 
	\begin{cases}
		\frac{ e^{\tau / \tau_\textrm{si}} }{\tau_\textrm{f} + \tau_\textrm{si}} &, \tau < 0   \\
		\frac{ e^{-\tau / \tau_\textrm{f}} }{\tau_\textrm{f} + \tau_\textrm{si}} +
		\frac{ e^{-\tau / \tau_\textrm{si}} -  e^{-\tau / \tau_\textrm{f}} }{\tau_\textrm{si} - \tau_\textrm{f}} &, \tau > 0 \,.
	\end{cases}	
\label{eq:fluorcorr}
\end{align}
The fit of the heralded correlation function $\textrm{g}^{(2)}_\textrm{fi}(\tau)$ in Fig.~\ref{fig:spdc}(c) matches the experimental data very well, yielding the time constants of $\tau_\textrm{si}=\SI[multi-part-units = single]{7.4\pm0.4}{\ns}$ and $\tau_\textrm{f}=\SI[multi-part-units = single]{37\pm1.1}{\ns}$. This corresponds to an extra-bandwidth added by the statistical interaction with the cesium atoms of $\Gamma_\textrm{meas}=\SI[multi-part-units = single]{4.3\pm0.3}{\MHz}$, in good agreement with the natural linewidth $\Gamma_\textrm{lit} \approx \SI{4.58}{\MHz}$ of the Cs D1 line. We measured a Klyshko efficiency of \SI{8.4}{\percent} for the idler photons and \SI{0.067}{\percent} for the re-emitted signal photons from  fluorescence. Here, an improved collection efficiency of the photons \cite{Maiwald2012} could enhance the Klyshko efficiency and lower the measurement time (see Fig.~\ref{fig:spdc}) both by orders of magnitude. 

The maximum of the correlation function from heralded fluorescence is delayed with respect to the reference measurement presented in Fig.~\ref{fig:spdc}(a) by $\Delta\tau_\textrm{} \approx \SI{10}{\ns}$ agreeing well with $\tau_\textrm{max}(\tau_\textrm{si},\tau_\textrm{f}) = \frac{\tau_\textrm{si} \cdot \tau_\textrm{f}}{\tau_\textrm{si} - \tau_\textrm{f}} \cdot  \textrm{ln}(2\,\tau_\textrm{si}/(\tau_\textrm{si}+\tau_\textrm{f}))$ given by our model. The observed delay is typical for the response of a damped oscillator driven by a pulse. A similar response is observed in optical resonators \cite{Bader2013}.
\section{Conclusion}
We demonstrated tunable and efficient interaction of two narrowband atomic resonances with photons created via spontaneous parametric down-conversion in a whispering-gallery mode resonator. Our system is not limited by the typical trade-off between the tunability and purity of the generated narrowband photons. The stringent phase-matching conditions of this triply-resonant cavity allow for single-mode operation without narrowband filtering. Going below the threshold of our optical parametric oscillations, we measured nearly complete absorption of the signal photons at the D1 lines of rubidium and cesium. The stability and brightness of our source greatly reduces the respective measurement times. This enables a direct probing of cesium D1 hyperfine levels with signal photons from fluorescence heralded by idler photons in a telecom band. To achieve this, we employed novel tuning mechanisms introduced with an evaluation of the mode spectrum and the frequency shift based on a movable dielectric substrate. Future applications of this source involve the generation of photon pairs and squeezed light in combination with resonant systems such as quantum dots, single atoms or optomechanical resonances. Multiphoton transitions are  easily accessible via the resonator-enhanced SPDC process. Further studies may also include different materials of the WGMR beyond lithium niobate to cover transitions with shorter wavelengths and study phase-matching configurations other than Type-I.
\section*{Acknowledgments}
The authors kindly acknowledge the support from Josef F\"{u}rst, Ralf Keding, Andrea Cavanna, and Felix Just.

\section*{Funding information}
Alexander von Humboldt Foundation; European 321
Research Council (ERC) (PACART).


\end{document}